\documentstyle[preprint,aps]{revtex}
\tightenlines

\begin{document}
\title{Possible peculiarities of synchrotron radiation in a strong magnetic field\footnote{%
Space Science and Technology (Kosmichna Nauka i Tehnologiya),
Kiev. Appendix. 2001, Vol.7, no 2, p.84--88}}
\author{B.I. Lev$^{1,2}$\footnote{%
Electronic addresses: lev@iop.kiev.ua, blev@i.kiev.ua}}
\author{A.A. Semenov$^1$\footnote{%
Electronic address: sem@iop.kiev.ua}}
\author{C.V. Usenko$^2$\footnote{%
Electronic addresses: usenko@phys.univ.kiev.ua,
usenko@ups.kiev.ua}}
\address{$^1$Institute of Physics, National Academy of Science of Ukraine. 46,\\
Nauky pr., Kiev 03028, Ukraine \\
$^2$Taras Shevchenko Kiev University, Physics Department. 6,\\
Academician Glushkov pr., Kiev 03127, Ukraine}
\date{\today }
\maketitle
\pacs{03.65.Bz, 03.65.Ca, 03.65.Sq.}

\begin{abstract}
Relativistic quantum effects on physical observables of scalar charged
particles are studied. Possible peculiarities of their behavior that can be
verified in an experiment can confirm several fundamental conceptions of
quantum mechanics. For observables independent of charge variable, we
propose relativistic Wigner function formalism that contains explicitly the
measurement device frame. This approach can provide the description of
charged particles gas (plasma). It differs from the traditional one but is
consistent with the Copenhagen interpretation of quantum mechanics. The
effects that are connected with this approach can be observed in
astrophysical objects - neutron stars.
\end{abstract}

\section{ Introduction}

The wave function nature has been considered as philosophical rather than
physical question for a long time. However it is very actual now because of
the recent theoretical and experimental progress in quantum information \cite
{12}.

In 1980, the specific behavior of the quantum systems that are
described by Einstein - Podolsky - Rosen (EPR) paradox was
confirmed in an experiment \cite{1}. It is very important that
EPR correlation ''spread'' in a space instantly. Nevertheless, if
one adheres to the Copenhagen interpretation there is no a
casualty principle breaking.

In a contrast if one can try to hold an objective and
deterministic description of quantum mechanics then classical
understanding of the casualty principle should be broken, because
of the conflict between quantum mechanics and special relativity
\cite{2,6}. In \cite{4} the relativistic classical and quantum
mechanics that generalizes the relativity principle was
constructed. Such an approach follows from the Eberhard and Bell
idea that correct description of quantum mechanics should contain
a preferred frame.

This theory has a well definite position operator. In the
preferred frame it coincides with the Newton - Wigner coordinate
\cite{7}. It means that measurement of the coordinate does not
create a particle-antiparticle couple because of there is no an
odd part.

Hence, it is very important to find situations when the odd part
of the position operator could manifest itself. But unfortunately
on the Earth, such experiments are very difficult due to a very
strong field needed. There are such fields near the astrophysical
objects - neutron stars. Therefore, it is interesting to find how
the odd part of the position operator can influence the
observable variables in a many-body system (in a gas of charged
particles or plasma).

The Wigner function formalism \cite{14} is a convenient method to describe
such systems. But there are several problems to generalize it to the
relativistic case. The first problem is that the time is not a dynamical
variable in the Weyl rule. In \cite{5} this problem was resolved by the
generalization of the spatial integration over the whole space-time. The
Wigner function formulation in a framework of the stochastic formulation of
quantum mechanics is Lorentz invariant too \cite{10}.

Formalism of the matrix-valued Wigner function for spin 1/2
particles was developed in \cite{8} with usual Weyl rule.
Certainly, such equations are not Lorentz invariant.

Next problem is the absence of well-defined position operator. In
\cite{13} the Wigner function formalism was developed using the
Newton - Wigner coordinate. The results in this approach differ
from the standard one. However one can connect they with \cite{4}
where the correctly definition of the position operator is
possible.

The aim of this work is a formulation of the Wigner formalism for scalar
charged particles in the approach \cite{8}. In addition we try to find
several peculiarities of the behavior of relativistic quantum system
including those ones with the complicated structure of the position operator.

\section{ Wigner function for a free particle}

To develop the Wigner function formalism one needs to formulate
the Weyl rule. Following \cite{8} should take into account that
classical variables are matrixes. But we restrict ourselves with
those proportional to the identity matrix. For a convenience we
shall use the Feshbach - Villars representation \cite{7}. The
Weyl rule is defined by usual way:
\[
\hat{A}_{\alpha }{}^{\beta }=\int {A(p,q)\hat{W}_{\alpha
}{}^{\beta }(p,q)dpdq,}
\]
where $\alpha ,\beta =\pm 1$ , $A(p,q)$ is the classical variable, $\hat{A}%
_{\alpha }{}^{\beta }$ is the corresponding classical variable, $\hat{W}%
_{\alpha }{}^{\beta }$ is the operator of quasi-probability
density that can be presented via the displacement operator:
\begin{equation}
\hat{W}_{\alpha }{}^{\beta }(p,q)=\frac{1}{{(2\pi \hbar )^{2d}}}\int {\hat{D}%
_{\alpha }{}^{\beta }(P,Q)e^{\frac{i}{\hbar }(Qp-Pq)}dQdP.}
\label{e1}
\end{equation}

In this representation the displacement operator can be expanded
by eigenvectors of the momentum operators:
\begin{equation}
\hat{D}_{\alpha }{}^{\beta }(P,Q)=\int {\left|
{p+\frac{P}{2}}\right\rangle
R_{\alpha }{}^{\beta }(p+\frac{P}{2},p-\frac{P}{2})e^{-\frac{i}{\hbar }%
Qp}dp\left\langle {p-\frac{P}{2}}\right| .}  \label{e2}
\end{equation}

In contrast to \cite{13} and non-relativistic case there is a
matrix-valued variable in (\ref{e2}):
\[
R_{\alpha }{}^{\beta }(p_{1},p_{2})=\varepsilon
(p_{1},p_{2})\delta _{\alpha }{}^{\beta }+\chi (p_{1},p_{2})\tau
_{1}{}_{\alpha }{}^{\beta }.
\]
It contains even and odd parts and is expressed via the relativistic energy
of a free particle $E(p)$ :

\begin{equation}
\begin{tabular}{l}
$\varepsilon (p_{1},p_{2})=\frac{{E(p_{1})+E(p_{2})}}{{2\sqrt{%
E(p_{1})E(p_{2})}}}$ \\
$\chi (p_{1},p_{2})=\frac{{E(p_{1})-E(p_{2})}}{{2\sqrt{E(p_{1})E(p_{2})}}}$%
\end{tabular}
.  \label{e3}
\end{equation}

Now combining (\ref{e1}) and (\ref{e2}) we obtain the formula for
the operator of quasi-probability density expansion:
\begin{equation}
\hat{W}_{\alpha }{}^{\beta }(p,q)=\frac{1}{{(2\pi \hbar )^{d}}}\int {\left| {p+%
\frac{P}{2}}\right\rangle R_{\alpha }{}^{\beta }(p+\frac{P}{2},p-\frac{P}{2}%
)e^{-\frac{i}{\hbar }Pq}dP\left\langle {p-\frac{P}{2}}\right| .}  \label{e4}
\end{equation}

The Wigner function is the average of this operator on an
arbitrary state:
\begin{equation}
W(p,q)=\sum\limits_{\alpha ,\beta }{\left\langle {\psi _{\beta }}\right| }%
\hat{W}_{\alpha }{}^{\beta }(p,q)\left| {\psi ^{\alpha
}}\right\rangle .\label{e5}
\end{equation}
This expression contains four terms. Two of them are the average of the even
part of the operator of quasi-probability density and two others are the
average of the odd part. Now one can introduce the symbols:
\begin{equation}
W_{\alpha }{}^{\beta }(p,q)=\left\langle {\psi _{\beta }}\right| \hat{W}%
_{\alpha }{}^{\beta }(p,q)\left| {\psi ^{\alpha }}\right\rangle .
\end{equation}
It should be noted that $W_{\alpha }{}^{\beta }(p,q)$ is not the
matrix-valued Wigner function in the sense \cite{8}.

Using the expressions (\ref{e4}) and (\ref{e5}) the components of
the Wigner function are obtained in the form:
\begin{equation}
\begin{tabular}{l}
$W_{\alpha }{}^{\alpha }(p,q)=\frac{1}{{(2\pi \hbar )^{d}}}\int
{\varepsilon (p+\frac{P}{2},p-\frac{P}{2})\psi _{\alpha }^{\ast
}(p+\frac{P}{2})\psi
^{\alpha }(p-\frac{P}{2})e^{-\frac{i}{\hbar }Pq}dP}$ \\
$W_{\alpha }{}^{-\alpha }(p,q)=\frac{1}{{(2\pi \hbar )^{d}}}\int {\chi (p+%
\frac{P}{2},p-\frac{P}{2})\psi _{\alpha }^{\ast }(p+\frac{P}{2})\psi
^{-\alpha }(p-\frac{P}{2})e^{-\frac{i}{\hbar }Pq}dP}$%
\end{tabular}
.
\end{equation}

One can obtain the quantum Liouville equation by standard way
\cite{14}:
\begin{equation}
\begin{tabular}{l}
$\frac{{\partial W_{\alpha }{}^{\alpha }(p,q,t)}}{{\partial t}}=\alpha \frac{2%
}{\hbar }E(p)\sin \{-\frac{\hbar }{2}\overleftarrow{\partial }_{p}%
\overrightarrow{\partial }_{q}\}W_{\alpha }{}^{\alpha }(p,q,t)$ \\
$\frac{{\partial W_{\alpha }{}^{-\alpha }(p,q,t)}}{{\partial t}}=i\alpha \frac{%
2}{\hbar }E(p)\cos \{\frac{\hbar }{2}\overleftarrow{\partial }_{p}%
\overrightarrow{\partial }_{q}\}W_{\alpha }{}^{-\alpha }(p,q,t)$%
\end{tabular}
.  \label{e6}
\end{equation}

The equation for the even part of the Wigner function coincides
with the similar expression in the Newton - Wigner coordinate
approach \cite{13}. Hence the dynamics of the distribution
function in both cases are identical. The difference is in the
constraints of the initial conditions. The physical variables
that contain higher moments of the coordinate (for example
dispersion) differs from those in \cite{13}. For one particle
problem these peculiarities were developed in \cite{3}.

\section{Particles in a homogeneous magnetic field}

The particles in external electromagnetic fields are more
sensitive to the odd part of the coordinate. For example, in a
uniform electric field the odd part of the position in the
Hamiltonian of interaction results in the effects of particles
creation from the vacuum \cite{9}. The origin of this peculiarity
is that even and odd parts of the Wigner function are entangled
in equations like (\ref{e6}).

Here we shall study the behavior of particles in a
time-independent and homogeneous magnetic field that is more
typical for astrophysical objects. Following to \cite{11} we
shall use the energy representation and so we have to consider
quasi-particles rather than particles. Both the position and
momentum operators have odd parts in this approach.

Further we do not take into account the particle motion along the
magnetic field and consider only relativistic rotator. Following
to the previous paragraph one can write the displacement operator
in the energy representation:
\begin{equation}
D_{n,m;}{}_{\alpha }{}^{\beta }(P,Q)=(\varepsilon _{n,m}\delta
_{\alpha }{}^{\beta }+\chi _{n,m}\tau _{1}{}_{\alpha }{}^{\beta
})D_{n,m}(P,Q),
\end{equation}
where $D_{n,m}(P,Q)$ are the matrix elements of the usual displacement
operator on the eigenfunctions of the harmonic oscillator \cite{15} and $%
\varepsilon _{n,m}$ , $\chi _{n,m}$ are defined like (\ref{e3})
but with the spectrum of the relativistic rotator in place of the
energy of a free particle. Then the operator of quasi-probability
density and the Wigner function are defined in the way presented
in the previous paragraph. The final expressions for the even and
odd components of the Wigner function are
\begin{equation}
\begin{tabular}{l}
$W_{\alpha }{}^{\alpha }(p,q)=\sum\limits_{m,n}{\varepsilon
_{m,n}C_{n;\alpha
}^{\ast }C_{m;}^{\alpha }{\rm T}_{m,n}(p,q)}$ \\
$W_{\alpha }{}^{-\alpha }(p,q)=\sum\limits_{m,n}{\chi
_{m,n}C_{n;\alpha
}^{\ast }C_{m;}^{-\alpha }{\rm T}_{m,n}(p,q)}$%
\end{tabular}
.
\end{equation}
Here $C_{n;\alpha }$ is the wave function in the energy representation, $%
T_{m,n}(p,q)$ is the matrix elements of the usual operator of
quasi-probability density
\begin{equation}
\hat{T}(p,q)=\frac{1}{{(2\pi \hbar )^{2d}}}\int {\hat{D}(P,Q)e^{-\frac{i}{%
\hbar }(Pq-Qp)}dPdQ}.
\end{equation}

The equations for the Wigner function (12) can be obtained in the
standard way too. Here the different components are not
entangled. Hence there are no any effects connected with vacuum
instability \cite{9}:
\begin{equation}
\begin{tabular}{l}
$\frac{{\partial W_{\alpha }{}^{\alpha }(p,q,t)}}{{\partial t}}=\alpha \frac{2%
}{\hbar }E(p,q)\sin \{\frac{\hbar }{2}(\overleftarrow{\partial }_{q}%
\overrightarrow{\partial }_{p}-\overleftarrow{\partial }_{p}\overrightarrow{%
\partial }_{q})\}W_{\alpha }{}^{\alpha }(p,q,t)$ \\
$\frac{{\partial W_{\alpha }{}^{-\alpha }(p,q,t)}}{{\partial t}}=i\alpha \frac{%
2}{\hbar }E(p,q)\cos \{-\frac{\hbar }{2}(\overleftarrow{\partial }_{q}%
\overrightarrow{\partial }_{p}-\overleftarrow{\partial }_{p}\overrightarrow{%
\partial }_{q})\}W_{\alpha }{}^{-\alpha }(p,q,t)$%
\end{tabular}
.  \label{e7}
\end{equation}
In this expression we introduce the Weyl symbol for the Hamiltonian of the
relativistic rotator in the energy representation (it should be redefined):
\[
E(p,q)=mc^{2}\sqrt[\star]{1+\frac{2}{{mc^{2}}}(\frac{{p^{2}}}{{2m}}+\frac{{\omega
_{c}^{2}m}}{2}q^{2})},
\]
where $\omega _{c}=\frac{{eB}}{m}$ is the cyclotron frequency, and
the square root is defined in the sense of star-product
$\star\equiv e^{\frac{i\hbar}{2}\left(\overleftarrow{\partial }_{q}%
\overrightarrow{\partial }_{p}-\overleftarrow{\partial }_{p}\overrightarrow{%
\partial }_{q}\right)}$ .

The odd part in the Wigner function definition describes
interference effects between particles and antiparticles.
Furthermore the value $\varepsilon _{nm}$ defines the
specification of the initial conditions.

In \cite{11} we studied dynamics of the one particle problem. It
was shown that in fields less than critical ($\hbar \omega
_{c}<mc^{2}$ ) the mean radius of the trajectory
oscillates with the frequency $\Omega =\frac{{\hbar \omega _{c}^{2}}}{{mc^{2}}%
}$ . This is essentially quantum relativistic effect. It can be observed as
a low frequency modulation of synchrotron radiation. But unfortunately such
dynamical process is identical for both approaches and does not reveal the
complicated structure of the position operator.

Now let us consider dispersion of the orbit radius for the
nonlinear coherent state \cite{11}:
\[
\overline{\Delta R^{2}}=a^2-\overline{R}^2\left[ {1-\left| {%
C_{\circ }}\right|
^{2}\sum\limits_{n}{\frac{{\overline{R}^{2n}}}{{2^{n}a^{2n}n![{\varepsilon
_{n+1,n+2}}^2]!}}}}\right] ,
\]
where $C_{\circ }$ is the normalization factor, $a^2=\frac{\hbar
}{{m\omega _{c}}}$. It contains both the standard and additional
terms. They result in the appearance of the states with formally
broken uncertainty relation. One can expect that such effects
take place for many-body systems too.

\section{Conclusion}

The odd part of the position operator results in the non-standard behavior
of the physical observables. The whole system has peculiarities too. However
it is observed not for all physical variables. For example, behavior of
energy does not contain such peculiarities. Hence one can expect these
effects for the quadratic and higher moments of the coordinate and momentum.

Especially one should notice the effects connected with
interference between particles and antiparticles. They result
from the odd part of the Wigner function and can be observed in
systems of particles with opposite charge signs.

Finally we briefly note that relativistic quantum mechanics in a
Wigner formulation contain the measurement device frame. Actually
one can write the equation (\ref{e6}) and (\ref{e7}) using only
four dimensional Lorentz invariant symbols. To make it possible
one should incorporate into equations certain time-like vector. It
can be interpreted as a four-velocity of the frame where a wave
packet reduction happens relatively to the second (immobile)
observer. It is very important that quantum mechanics equations
contain explicitly the observer characteristics. This fact can
serve as an additional argument in favor of the Copenhagen
interpretations of quantum mechanics.


\begin{references}
\bibitem{1}  Aspect A., Roger G., et. al. Time Correlations between Two
Sidebands of the Resonance Fluorescence Triplet // Phys.Rev.Lett. 1980, 45,
p. 617; Aspect A., Grangier P., Roger G. Experimental Test of Realistic
Local Theories via Bell's Theorem // Phys. Rev .Lett. 1981, 47, p.460.

\bibitem{2}  Bell J.S. Speakable and Unspeakable in Quantum Mechanics.
Cambridge: Cambridge University Press, 1987.

\bibitem{3}  Blokhintzev D.I.. On localization relativistic micro-particles
in space and time. JINR, R-2631, Dubna, 1966.[in Russian]

\bibitem{4}  Caban P., Rembielinski J. Lorentz-covariant quantum mechanics
and preferred frame // Phys. Rev. A. 1999, 59, p.4187.

\bibitem{5}  De Groot S.R., van Leeuwen W.A., Weert Ch.G. Relativistic
kinetic theory. Principles and Applications. Amsterdam: ''North-Holland'',
1980.

\bibitem{6}  Eberhard P.H. Bell's Theorem and the Different Concepts of
Locality. // Nuov. Cim. 1978, 46B, 392.

\bibitem{7}  Feshbach H., Villars F. Elementary Relativistic Wave Mechanics
of Spin 0 and Spin 1/2 Particles // Rev. Mod. Phys. 1958, 30, p.24.

\bibitem{8}  Gerard P., Markovich P.A., Mauser N.J. et al. Homogenization
Limits and Wigner Transforms // Comm. Pure Appl. Math. 1997, 50, p.0323.

\bibitem{9}  Grib A.A., Mamaev S.G., Mostapenko V.M. Vacuum quantum effects
in strong fields. Moscow: ''Energoatomizdat'', 1988.[in Russian]

\bibitem{10}  Holland P.R., Kyprianidis A., Maric Z. et al. Relativistic
generalization of the Wigner function and its interpretation in the casual
stochastic formulation of quantum mechanics // Phys. Rev. A. 1986, 33,
p.4380.

\bibitem{11}  Lev B.I., Semenov A.A., Usenko C.V. Behaviour of $\pi ^{\pm }$
mesons and synchrotron radiation in a strong magnetic field. // Phys. Lett.
A. 1997, 230, p.261.

\bibitem{12}  Menskii M.B. Quantum mechanics: new experiments, new
applications, new formulations // Uspekhi - Physics. 2000, 43, 6, p.585.

\bibitem{13}  Mourad J. The Wigner-Weyl formalism and the relativistic
semi-classical approximation // Phys. Lett. A. 1993, 179, p. 231. E-print of
LANL: /hep-th/9307135.

\bibitem{14}  Moyal J.E. Quantum Mechanics as a Statistical Theory // Proc.
Cambr. Phil. Soc. 1949, 45, p.99.

\bibitem{15}  Perelomov A.M. Generalized coherent states and their
applications. Berlin: ''Springer'', 1996.
\end{references}
\end{document}